\documentclass[aps,twocolumn,amsmath,amssymb,prl,aps]{revtex4-1}

\usepackage{graphicx}
\usepackage{dcolumn}
\usepackage{amssymb}
 \usepackage{amsthm}
 \usepackage{amsmath}
 \usepackage{color}
 
 \usepackage{lipsum}  
 \usepackage{soul} 
 
\begin{document}


\title{Adaptive Hagen-Poiseuille flows on graphs}


\author{Rodrigo Almeida}

\author{Rui Dil\~ao}
\email[]{ruidilao@tecnico.ulisboa.pt}
\affiliation{University of Lisbon, Instituto Superior T\'ecnico, Nonlinear Dynamics Group\\
Av. Rovisco Pais, 1049-001 Lisbon, Portugal}

\date{\today}

\begin{abstract}
We derive a class of equations describing  low Reynolds number steady flows of  incompressible and viscous fluids in networks made of straight channels, with several sources and sinks, and adaptive conductivities. The flow is controlled by the fluxes at sources and sinks. 
The network is represented by a graph and the adaptive conductivities describe the transverse channel elasticities, mirroring several network structures found in physics and biology.  Minimising the dissipated energy per unit time, we have found an explicit form for the adaptation equations and, asymptotically in time, a steady state tree geometry for the graph connecting sources and sinks is reached. 
A phase transition tuned by an order parameter for the adapted steady sate graph has been found.
\end{abstract}


\maketitle


Systems with network structures characterizing their functioning are veins in tumours \cite{Rub}, channels transporting nutrients in leaves \cite{Kat, Ron},   animals' vasculature \cite{Hu}, microorganisms with adaptive vein networks as the \textit{Physarum polycephalum} \cite{Bau, Oet},  and \textit{Dictyostelium discoideum} \cite{Alm},  transportation systems  \cite{Ter2}, or electrical networks \cite{Bon}. Formal systems  as the travelling salesman problem \cite{Zhu}, the Steiner tree geometries \cite{Ter5},  and the first passage percolation \cite{How}  are challenging network optimisation problems in graphs, with important applications in physics, biology and engineering.

\textit{Physarum polycephalum} is a slime mould with a complex life cycle, showing a great adaptability to environmental changes. It has an adaptive network of veins, which transport the endoplasmatic fluid from food sources to the different regions of the slime mould body, and whose geometry and adaptability are responsible for growth and motility \cite{Ali1}. The veins walls of the network show rhythmic transverse oscillations,   regulated by food sources and other external stimuli, \cite{Ali2} and \cite{Ali3}.  Despite lacking any kind of neural circuit, \textit{Physarum} shows a high-level behaviour, where, for example, the choice of the shortest path between food sources in a maze has been observed \cite{Ter3}. Moreover, 
in the presence of multiple food sources, \textit{Physarum}  can build networks  with  a  great  compromise  between  production cost, transport efficiency and fault tolerance
\cite{Ter4}. 
The main biochemical and physical mechanisms responsible by the rhythmic behaviour and the movements of \textit{Physarum} is a challenging problem and has not been identified \cite{Tep}.

In the particular case of the dynamics and optimisation of vein networks in \textit{Physarum}, several modelling approaches aiming to explore the formation and optimisation of network paths have been introduced,   \cite{Ter4}. These approaches were based on the phenomenology of Hagen-Poiseuille   flows in graphs,  and optimisation has been introduced by controlling channel conductivity through   \textit{ad hoc} local evolution laws, together with stochastic actualisation processes.  These models converge to steady   network paths, which, in some cases,  are close to Steiner tree type of connections, \cite{Ter5} and \cite{Cal}, or even shortest path solutions \cite{Ter3, Zhu}.

However, in all these modelling approaches some basic physical principles are not taken into account, as is the case of the conservation of  volume of the circulating fluid. On the other hand, it is difficult to understand that the local conductivity of channels could change independently of the other channel conductivities of the network, a hypothesis subjacent to  the modelling approaches just described.

The main goal of this paper is to construct a class of models describing the flow of incompressible and viscous fluids in networks of veins or channels, in conformity with the properties of fluids. The network is described by a graph where edges represent channels, and vertices are spatial distribution hubs of fluid. The  veins are  elastic and their diameters can change in response to the flux of fluid in the  channels. 


We consider simple undirected graphs ${\cal{G}}=(V,E)$, embedded in a  $n$ dimensional Euclidean space, where $V$ is the set of $N$ nodes or vertices with coordinates $(x_i^1,\ldots, x_i^n)$, with $i=1,\ldots, N$, and $E$ is the set of $M$ straight edges $(i,j)$, connecting nodes $i$ and $j$.  The graph ${\cal G}$ is connected in the sense that it has only one component. The  graph ${\cal G}$ describes the geometry of a network of veins.

Edges represent elastic cylindric channels that can expand or contract transversely to the flow. The edge connecting node $i$ to node $j$ has fixed length $L_{ij}$ and radius $r_{ij}$. The fluid in the network is viscous and incompressible, the flow is laminar (low Reynolds numbers) with no-slip boundary conditions, being a  Hagen-Poiseuille (H-P) type flow  \cite{Lan}. The dynamic viscosity of the fluid is $\eta$.

Networks have $K$ sources and $R$  sinks, located at fixed nodes. 
The flux of fluid from outside to the interior of the network and vice versa, occurring at node  $j$, is represented by $S_j$. If the  network has a source at node $j$, then $S_j>0$. If it has a sink at node $j$, then $S_j<0$, otherwise $ S_j=0$. As the fluid in the network of veins is incompressible, we have
\begin{equation}\displaystyle
\sum_{j=1}^NS_j=\sum_{j:\hbox{sources}}S_j+\sum_{j:\hbox{sinks}}S_j=0,
\label{eq1}
\end{equation}
and the volume of fluid in the network is  constant.

The steady state of the flow is characterised by the pressure $p_i$ at the node $i$, and the flux along the channel or edge connecting nodes $i$ and $j$ is 
\begin{equation}\displaystyle
Q_{ij}=\frac{\pi r_{ij}^4}{8\eta}\frac{(p_i-p_j)}{L_{ij}}:=D_{ij}\frac{(p_i-p_j)}{L_{ij}},
\label{eq2}
\end{equation}
where $D_{ij}$ is the conductivity of the channel $(i,j)$, and $D_{ij}=D_{ji}$. If $Q_{ij}>0$, then the flow is from $i$ to $j$. If $Q_{ij}<0$, the flow is from $j$ to $i$.  The ratios $D_{ij}/L_{ij}$ have the role of graph weights.

The volume of a channel connecting node $i$ to node $j$ is ${\cal V}_{ij}=\pi r_{ij}^2L_{ij}$. As $D_{ij}={\pi r_{ij}^4}/{8\eta}$, we have ${\cal V}_{ij}=\sqrt{8\pi\eta}L_{ij}\sqrt{D_{ij}}$. As the fluid is incompressible, the volume of fluid in the network is constant and  is
\begin{equation}\displaystyle
{\cal V}=\sum_{(i,j)\in E}{\cal V}_{ij}=\beta\sum_{(i,j)\in E}L_{ij}\sqrt{D_{ij}} ,
\label{eq3}
\end{equation}
where $\beta=\sqrt{8\pi\eta}$.

The steady state of the  H-P  flow in the network   is determined by the Kirchhoff law
\begin{equation}\displaystyle
\sum_{j: (i,j)\in E}Q_{ij}=S_i,\ i=1\ldots , N.
\label{eq4}
\end{equation}
The system of conservation laws \eqref{eq4} can be written in the form $Wp=S$, where, by \eqref{eq2}, $W$ is the generalised Laplacian matrix of the graph ${\cal G}$,  depending of $D_{ij}$ and $L_{ij}$, and $W=W^T$, \cite{God}. The  vector $S$ describes the sources and the sinks, obeys   condition  \eqref{eq1}, and $p$ is the vector of pressures. If $i\not=j$ and $(i,j) \in E$, then $w_{ij}<0$.  If $i\not=j$ and $(i,j) \not\in E$, then $w_{ij}=0$. As $w_{ii}>0$, the matrix $W$ is diagonally dominated. 
As the graph ${\cal G}$ has one component, then $\hbox{rank}(W)=N-1$, \cite{Gal}. 
As  $W{\bf 1}=0$, the eigenvector ${\bf 1}$ of $W$  corresponds to the zero eigenvalue with multiplicity $1$.
 Then, the system of equations \eqref{eq4} can be solved for all the pressures, up to an additive constant, and the fluxes are well determined by \eqref{eq2}.

The dissipated energy per unit time of the H-P steady flow is
\begin{equation}\displaystyle
{\cal P}=\sum_{(i,j)\in E}\frac{Q_{i,j}^2}{D_{ij}}L_{ij}.
\label{eq5}
\end{equation}

The steady state solution of the linear equations \eqref{eq4}, with sources and sinks, is obtained for fixed  conductivities $D_{ij}$ and lengths $L_{ij}$. However, as it has been observed in organisms, the flow in veins induce the increase or decrease  the radius of channels,  affecting the conductivity of veins, \cite{Rub2} and \cite{Woh}. Due to the incompressibility of the fluid, this affects all the network structure which can not be described by local mechanisms.

To obtain a law for the rate of change of the conductivities of the network, we assume  that the area of channels can change as a function of fluxes, and make the ansatz 
\begin{equation}\displaystyle
\frac{d}{dt}\sqrt{D_{ij}}=f({\bf Q})-c \sqrt{D_{ij}}\ ,\ (i,j)\in E
\label{eq6}
\end{equation}
where, by \eqref{eq2}, in a H-P flow, $\sqrt{D_{ij}}$ is proportional to the area of the $(i,j)$ channel,  $f({\bf Q})$,  with $f({\bf 0})=0$, is an unknown function of the global flux ${\bf Q}$ in the network, and $c >0$ is some constant.

Assuming conservation of volume,  by \eqref{eq3}, we have
\begin{equation}\displaystyle
\frac{d}{dt}{\cal V}=\beta\sum_{(i,j)\in E}L_{ij}\frac{d}{dt}\sqrt{D_{ij}}=0,
\label{eq7}
\end{equation}
which, by  \eqref{eq6} and \eqref{eq3}, implies that
\begin{equation}\displaystyle
\sum_{(i,j)\in E}L_{ij} f({\bf Q})=c \sum_{(i,j)\in E} L_{ij} \sqrt{D_{ij}}=\frac{c}{\beta} {\cal V}
\label{eq8}
\end{equation}
is constant.

Defining a new function $g$ by the relation
$$\displaystyle
f({\bf Q}):= \frac{c}{\beta} {\cal V} \frac{g(Q_{ij})}{\sum_{(k,m)\in E}L_{km}g(Q_{km})}
$$ 
and introducing this last expression into \eqref{eq6}, we conclude that
\begin{equation}\displaystyle
\frac{d}{d\tau}\sqrt{D_{ij}}=  \alpha   \frac{g(Q_{ij})}{\sum_{(k,m)\in E}L_{km}g(Q_{km})}- \sqrt{D_{ij}},\ (i,j)\in E
\label{eq9}
\end{equation}
where $\tau=c t$ and $\alpha={\cal V}/\beta$, and the first term on the right-hand side of the above equation depends on the fluxes of all the channels in the network. 
Therefore, we have shown that, for any choice of the function $g$,
the adaptation equation  \eqref{eq9} maintains invariant the volume of fluid in the network.  

To analyse the temporal evolution of the geometry of the network of veins and eventually its dynamical adaptation, given prescribed sources and sinks fluxes, the function $g$ in \eqref{eq9} must be chosen. For that, we introduce the simple criterion of minimisation of the dissipated energy per unit time during the flow 
(\cite{Fey, Van, Bon, Cor}), also preserving the total volume ${\cal V}$ of fluid. So,  by  \eqref{eq3} and  \eqref{eq5}, we consider the  Lagrangian 
$$
{\cal L}={\cal P}-\lambda \left({\cal V}-\beta\sum_{(k,m)\in E}L_{km}\sqrt{D_{km}} \right),
$$
where $\lambda$ is a Lagrange multiplier. Minimising ${\cal L}$ with respect to  $\sqrt{D_{ij}}$ and $\lambda$, by \eqref{eq5}, we obtain
$$
\left\{
\begin{array}{lcl}\displaystyle
\frac{\partial {\cal L}}{\partial \sqrt{D_{ij}}}&= &\displaystyle \left(-\frac{Q_{ij}^2}{D_{ij}^2}L_{ij}+2\sum_{(k,m)\in E} \frac{Q_{km}}{D_{km}}\frac{\partial Q_{km}}{\partial D_{ij}} L_{km}\right.\\
&&\displaystyle\left. +\lambda \beta \frac{L_{ij}}{2\sqrt{D_{ij}}}\right)2\sqrt{D_{ij}}=0\\ \displaystyle
\frac{\partial {\cal L}}{\partial \lambda}&=& \displaystyle {\cal V}-\beta\sum_{(k,m)\in E}L_{km}\sqrt{D_{km}}=0.
\end{array}\right.
$$
As the second term on the right-hand side of the first equation is zero (cf. lemma~2.1 in \cite{Has}), solving  these equations with respect to $D_{ij}$ and $\lambda$, the values of conductivities that minimize the dissipated energy per unit time of the flow are
\begin{equation}\displaystyle
D_{ij}^*=  \alpha^2  \frac{Q_{ij}^{4/3}}{\left(\sum_{(k,m)\in E}L_{km} Q_{km}^{2/3}\right)^2} 
\label{eq10}
\end{equation}
and $D_{ij}^{**}= 0$.

Comparing the conductivity solution \eqref{eq10}  with \eqref{eq9}, 
making the choice $g(Q_{ij})=Q_{ij}^{2/3}$ in \eqref{eq9}, by the  conditions $\frac{d}{d\tau}\sqrt{D_{ij}}= 0$, the dissipated energy per unit time at the steady state of the adaptive H-P flow is minimal.

For a single channel with one source and one sink, equation \eqref{eq9} reduces to $\frac{d}{d\tau}\sqrt{D_{12}}=\alpha/L_{12}-\sqrt{D_{12}}$, which has a unique stable fixe point for $\sqrt{D_{12}}=\alpha/L_{12}$, coinciding with \eqref{eq10}. Therefore, for a simple H-P flow in a channel and 
the conductivity  $D_{12}=\alpha^2/L_{12}^2=\pi r_{12}^4/\beta^2$, the dissipated energy per unit time of the H-P flow is minimal,  is independent of the choice of the function $g(.)$ and is consistent with the Kirchhoff law (cf. \cite{Fey} for this discussion). This property justifies the ansatz made in  \eqref{eq6} by adding the term $\sqrt{D_{ij}}$.

In the following, we analyse  volume conserving H-P flows in networks with adaptation equations
\begin{equation}\displaystyle
\frac{d}{d\tau}\sqrt{D_{ij}}=  \alpha   \frac{|Q_{ij}|^{\gamma}}{\sum_{(k,m)\in E}L_{km}|Q_{km}|^{\gamma}}- \sqrt{D_{ij}},\ (i,j)\in E
\label{eq11}
\end{equation}
where $\gamma >0$ is a new parameter  and the absolute value in $Q_{ij}$ emphasises that  conductivities are independent of the direction of the flow. 

The choice of a monotonous increasing function for $g(Q_{ij})$ may  lead to high conductivity values, only limited by the finitude of the graph ${\cal G}$ and by the volume of fluid in the network. Other choices for $g(Q_{ij})$ are possible, in particular those leading to saturation effects, \cite{Rub2} and \cite{Ter2}.
 
To analyse the flow of a fluid on a network and the adaptation of the elastic channels, we consider the simplest case where  
the network is described by a planar graph embedded in the two-dimensional Euclidean space, resulting from a Delaunay triangulation, and with nodes  in general position. Some of these nodes can be sources or sinks, with global fluxes obeying condition \eqref{eq1}. By increasing the number of nodes of the graph in a finite region of the Euclidean space, the advantage of this choice for the geometry of the graph  ${\cal G}$ is  the possibility of approaching in a simple way the  continuum limit  avoiding lattice symmetries.

To describe the flow on ${\cal G}$, we consider that  at some initial time $\tau=0$, all the edges of the network have positive conductivities and the flow is driven by the fluxes at sources and sinks. By \eqref{eq3}, the volume of the fluid in the network is determined by the initial conductivities $D_{ij}(0)$. The temporal evolution of the radius of the channels  starts with the evaluation of a H-P steady state by solving the linear equations \eqref{eq4}, for given initial conductivities $D_{ij}(0)$. In a second step, the conductivities of all the channels are adapted according to equations \eqref{eq11}. Then,  these two steps are repeated over time, until a steady state  of the channel conductivities is eventually reached. From the numerical point of view, a  steady state is reached if the condition  $\max_{(i,j)\in E}|D_{ij}(n\Delta\tau)-D_{ij}((n-1)\Delta\tau)|<10^{-6}$  is fulfilled, where $n$ is the first integer for which the inequality is verified and $\Delta\tau$ is the time increment to solve numerically equations \eqref{eq11}.   The time of convergence of the adaptation algorithm is $\tau^*=n\Delta \tau <\infty$.  
Several numerical experiments have shown that the above convergence criterium is a good compromise between computing time and the shape of the steady state channels.
According  to  equation \eqref{eq11}, the conductivities of edges may asymptotically approach  zero,  but  in the numerical simulations the limit is never reached,  and  thus,    the graph G remains always as a single connected component.

\begin{figure}[htb]
\centering
\includegraphics[width=0.49\columnwidth]{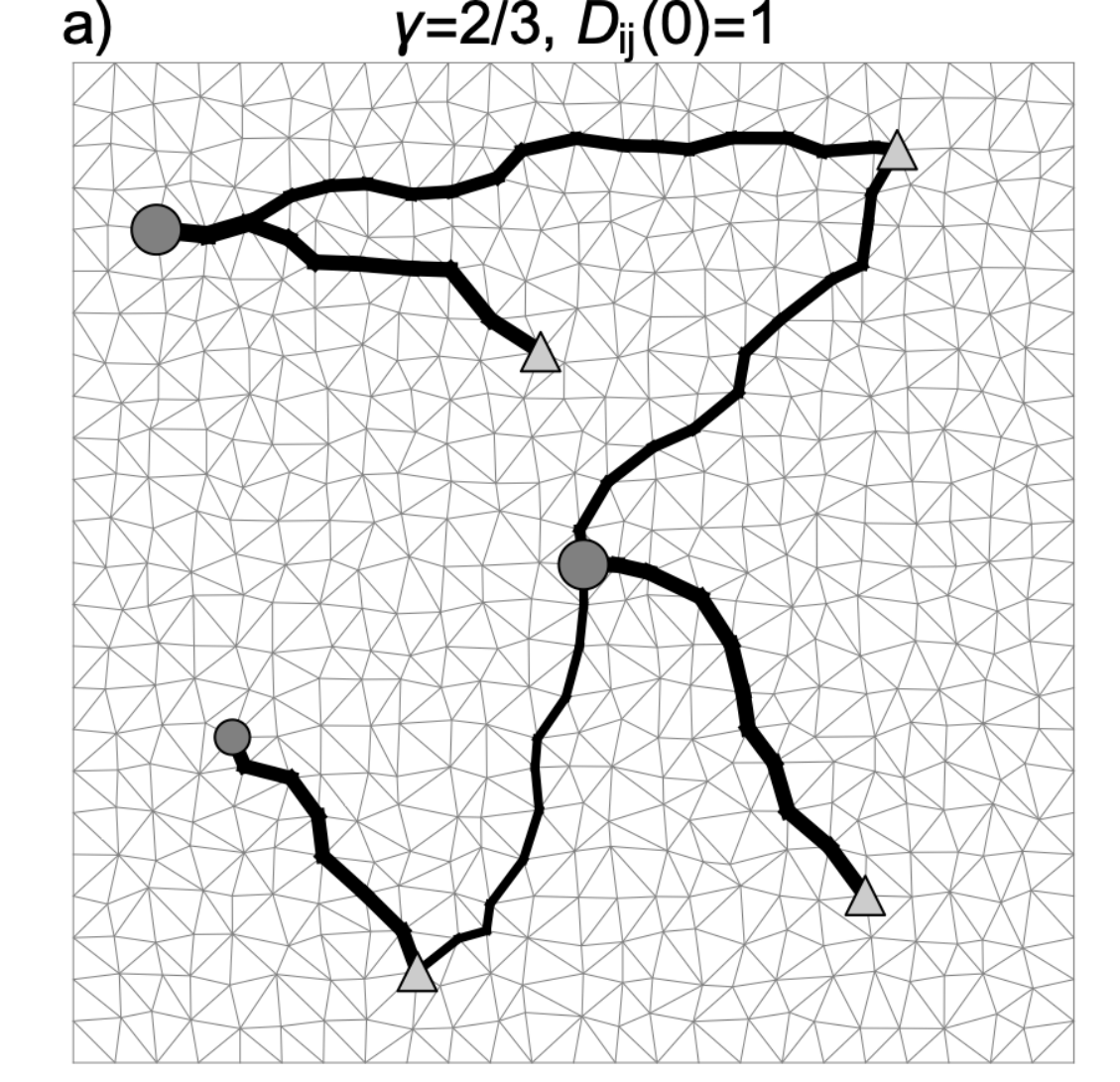}
\includegraphics[width=0.49\columnwidth]{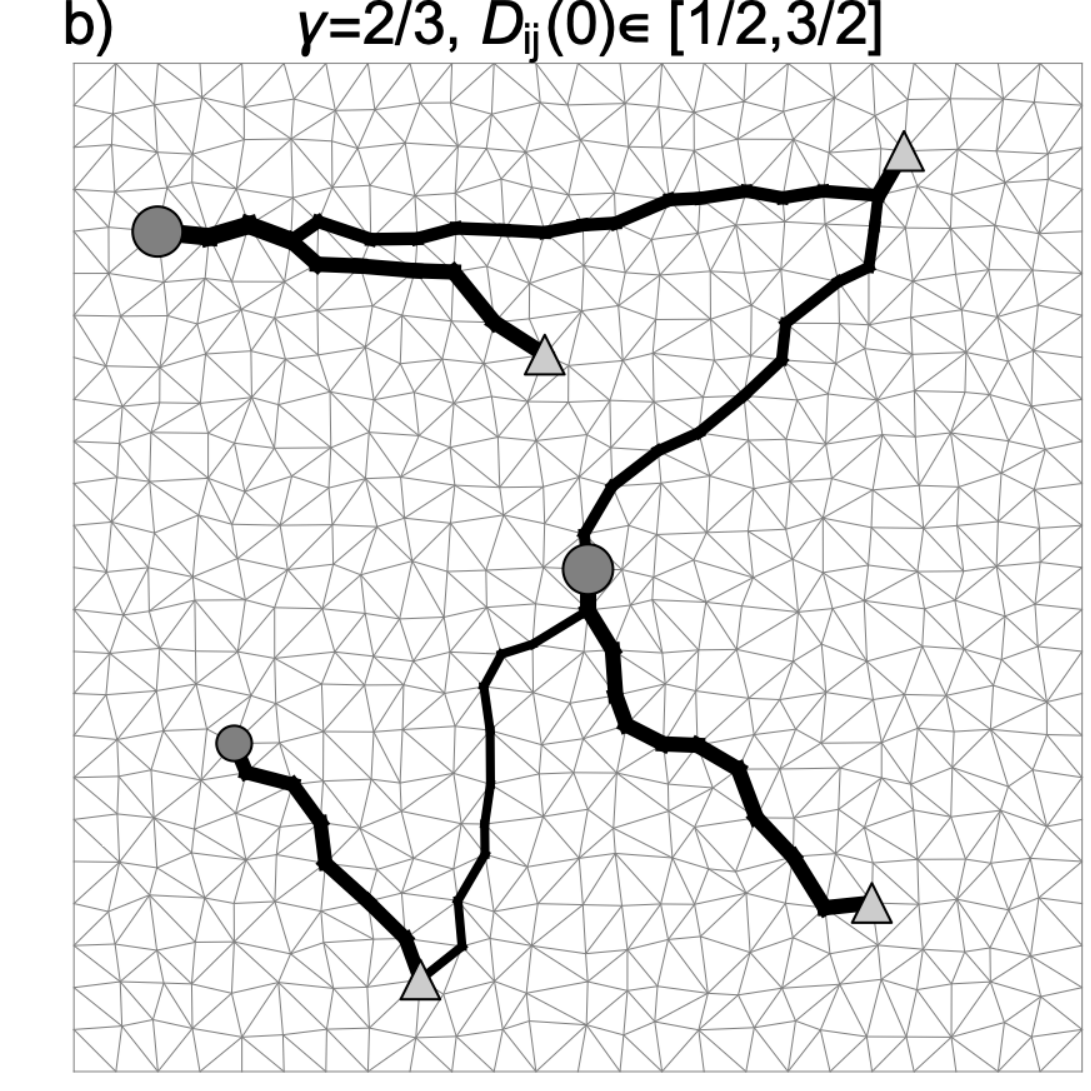}
 \caption{Steady states of an adaptive H-P flow with $3$ sources (circles) and $4$ sinks (triangles), calculated for two different initial distributions of conductivities, with adaptation calculated by equations \eqref{eq11}. The initial Delaunay triangulation ${\cal G}$ is shown in grey. 
The thickness of the black lines is proportional to the radius  of  the  edges ($\sim  D_{ij}^{1/4}$).
 At steady states, the dissipated energy per unit time is minimal.  The steady states solutions have been reached after $n=329$ iterations in a) and, after $n=444$ iterations in b). At steady state, $\tau^*=n\Delta \tau$, the tree conductivities obey to    $130<D_{ij}(\tau^*)<2750$, and the grey edges conductivities obey to  $0<D_{ij}(\tau^*)<0.0005$.}
\label{fig1}
\end{figure}

To test the adaptation mechanism \eqref{eq11}, we have considered a  square lattice of side length $1$, with  $N=25\times 25$ nodes. The positions of the interior nodes of the lattice have been randomly perturbed with a Gaussian noise, with standard deviation  $\sigma=0.5$. Then, a Delaunay triangulation of the lattice sites is performed, resulting the graph network ${\cal G}$. The  edge lengths  $L_{ij}$ are  then obtained.
We have considered $K=3$ sources and $R=4$ sinks, randomly allocated to the nodes of the network ${\cal G}$. Two sources have intensities $S_{source}=0.4$ and a third source has intensity  $S_{source}=0.2$. The sinks have equal intensities $S_{sink}=-0.25$. The adaptation process has been calculated with  equations \eqref{eq11}, for the  parameter values $\beta =1$ and $\gamma=2/3$, using the Euler explicit numerical method with time increment $\Delta\tau=0.1$.  (A simple calculation shows that the discretised Euler method preserves the volume of the channels in the graph, simplifying numerical calculations.) Two different sets of initial conductivities have been considered: homogeneous conductivities all over the edges, $D_{ij}(0)=1$, and random conductivities uniformly distributed in the interval $[1/2,3/2]$. In Figure~\ref{fig1}, we show the resulting adapted steady state networks. The steady state graphs in the figure are different but can be considered metrically close.  The lengths of the steady state trees are, in a), $L=2.936$, and, in b), $L=2.859$. For comparison, an upper bound for the length of the Steiner tree connecting sources and sinks is $L_{us}=2.333$, \cite{Kou}.

Several numerical simulations have shown that the steady state optimised network depends on the time step $\Delta\tau$, on the fluxes imposed by the sources and sinks, and on the initial distribution of conductivities.

If the number of sources and sinks are equal and the absolute value of their fluxes are the same, then the steady state may be a set of ``apparently" disconnected graph trees.   With a careful manipulation of the localisation of sources and sinks, it is possible to build networks with several ``apparently"  disconnected trees as steady states of the H-P flow.   In fact, for every $\tau\ge 0$, the  graph ${\cal G}$ has   only one connected component  with conductivities partitioned in  two sets of low and high values. 
This distribution of conductivities enables to count the number of effectively conducting channels of the steady state flow.
 
\begin{figure}[htb]
\centering
\includegraphics[width=0.5\columnwidth]{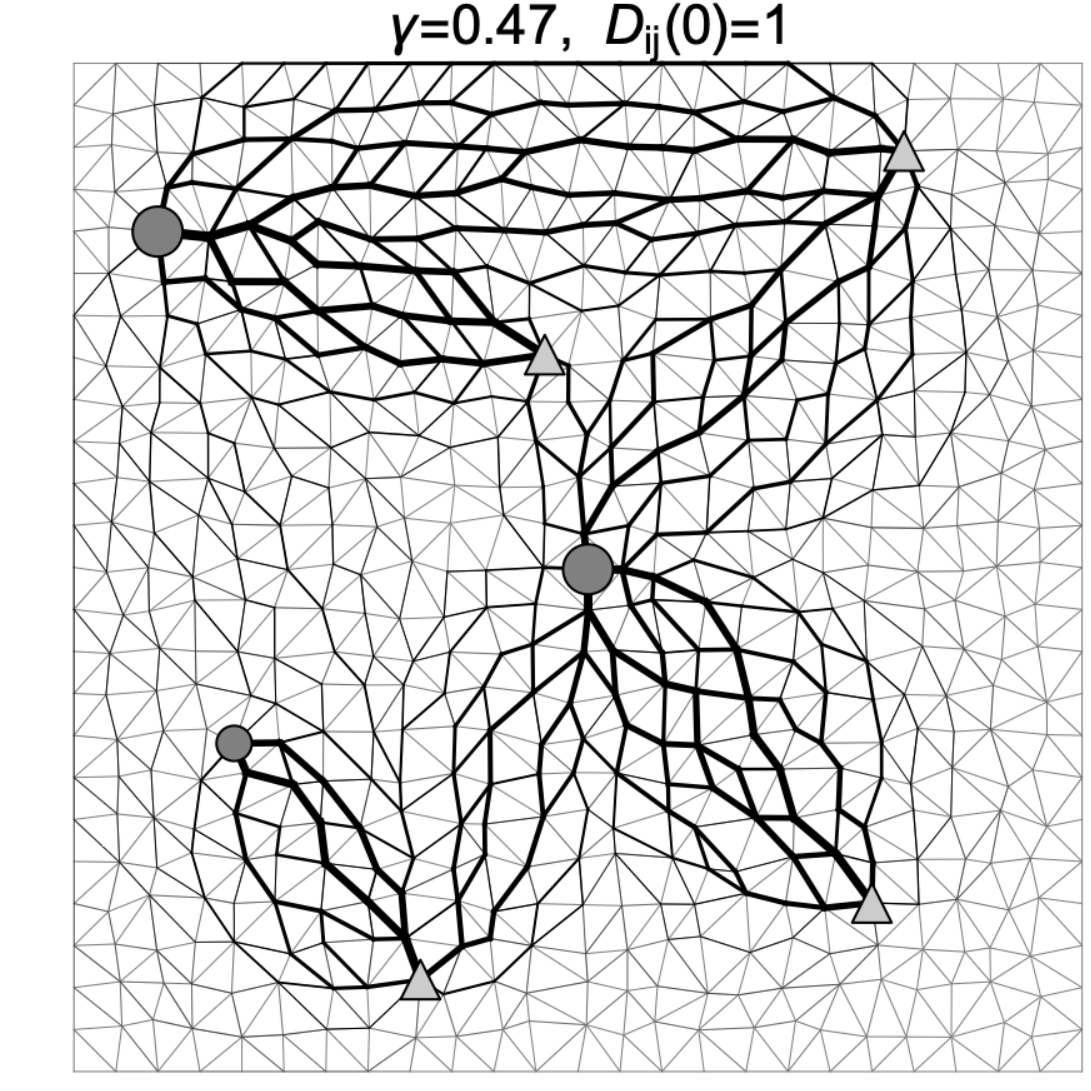}
 \caption{Steady state of the adaptive H-P flow for the order parameter $\gamma=0.47$, below the phase transition at $\gamma\simeq 1/2$, for the   initial network of Figure~\ref{fig1}a). The steady state network length is $L\simeq 41.50$, calculated at the iteration number  $n=2449$. The thickness of the black lines is proportional to the radius  of  the  edges.}
\label{fig2}
\end{figure}

Further numerical analysis of the adaptation equation \eqref{eq11} shows that a phase transition 
occurs for $\gamma\simeq 1/2$. If $\gamma <1/2$, the steady state of the  flow is a connected graph with closed loops, as shown in Figure~\ref{fig2}. As shown previously, if $\gamma >1/2$,  the steady state of the  flow  with meaningful conductivities is a tree connecting sources and sinks (Figure~\ref{fig1}).
This phase transition is consistent with the one found by Bohn and Magnasco \cite{Bon} in resistors network which minimizes the dissipation rate under the
cost constraint $\sum_{(i,j)}C_{ij}^\Gamma=\hbox{const}$,   where $C_{ij}$ are the edge
conductances. The conductances and the currents of the optimal network obey to a similar power-law relationship, with the natural identification of $\gamma = 1/(1 +\Gamma)$ (cf. equation (5) of \cite{Bon} with the first term on the right hand side of \eqref{eq11}). Using a relaxation algorithm to find the optimal network for each $\Gamma$, the authors have shown that the system exhibited a phase transition at $\Gamma=1$, with redundant uniform networks for $\Gamma>1$ and loopless trees for $\Gamma<1$.

\begin{figure}[htb]
\centering
\includegraphics[width=\columnwidth]{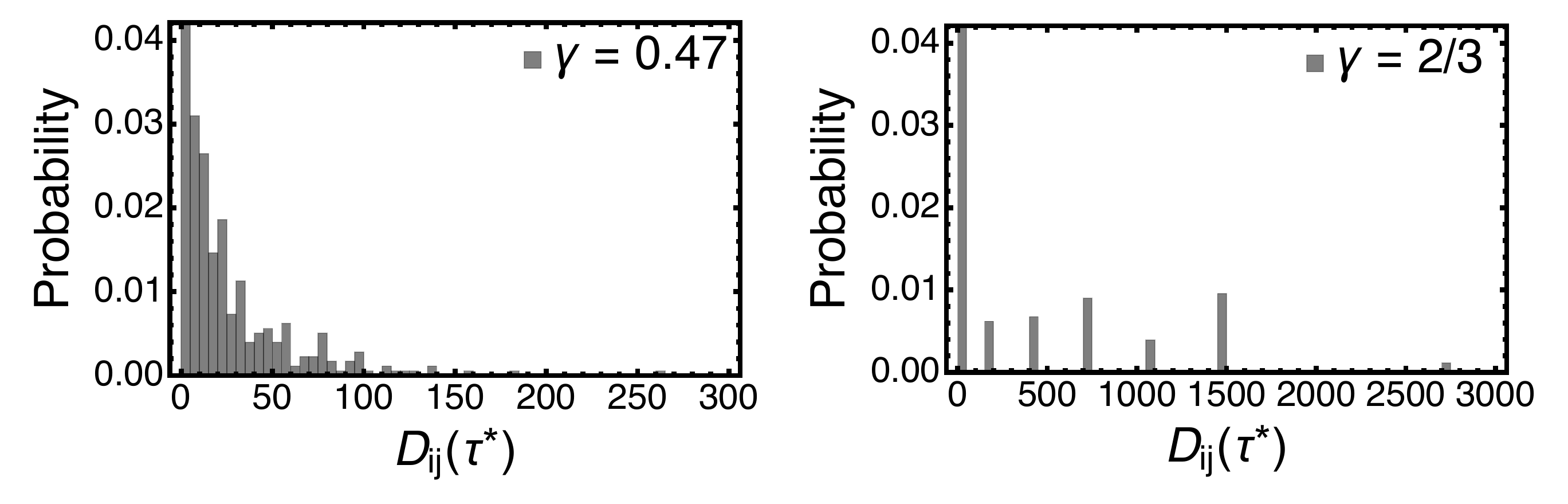}
 \caption{Histogram of frequencies of the steady state conductivities $D_{ij}(\tau^*)$,   for $\gamma=0.47$ and $\gamma=2/3$, and initial conditions as in Figure~\ref{fig1}a).}
\label{fig3}
\end{figure}

In Figure~\ref{fig3}, we show the histogram of frequencies of the steady state  conductivities $D_{ij}(\tau^*)$,  for $\gamma=0.47$ and $\gamma=2/3$,  and  initial conditions as in Figure~\ref{fig1}a). The distribution of conductivities enables the evaluation of the  steady state  network length $L$ by considering the edges of the full graph ${\cal G}$ with  conductivities above the threshold $D_{thr}=5\times 10^{-4}$.

The phase transition at $\gamma\simeq 1/2$ is quantified by analysing the dissipated energy per unit time at steady state, or by the  length of the steady state graph, as a function  of $\gamma $. 

\begin{figure}[htb]
\centering
\includegraphics[width=0.49\columnwidth]{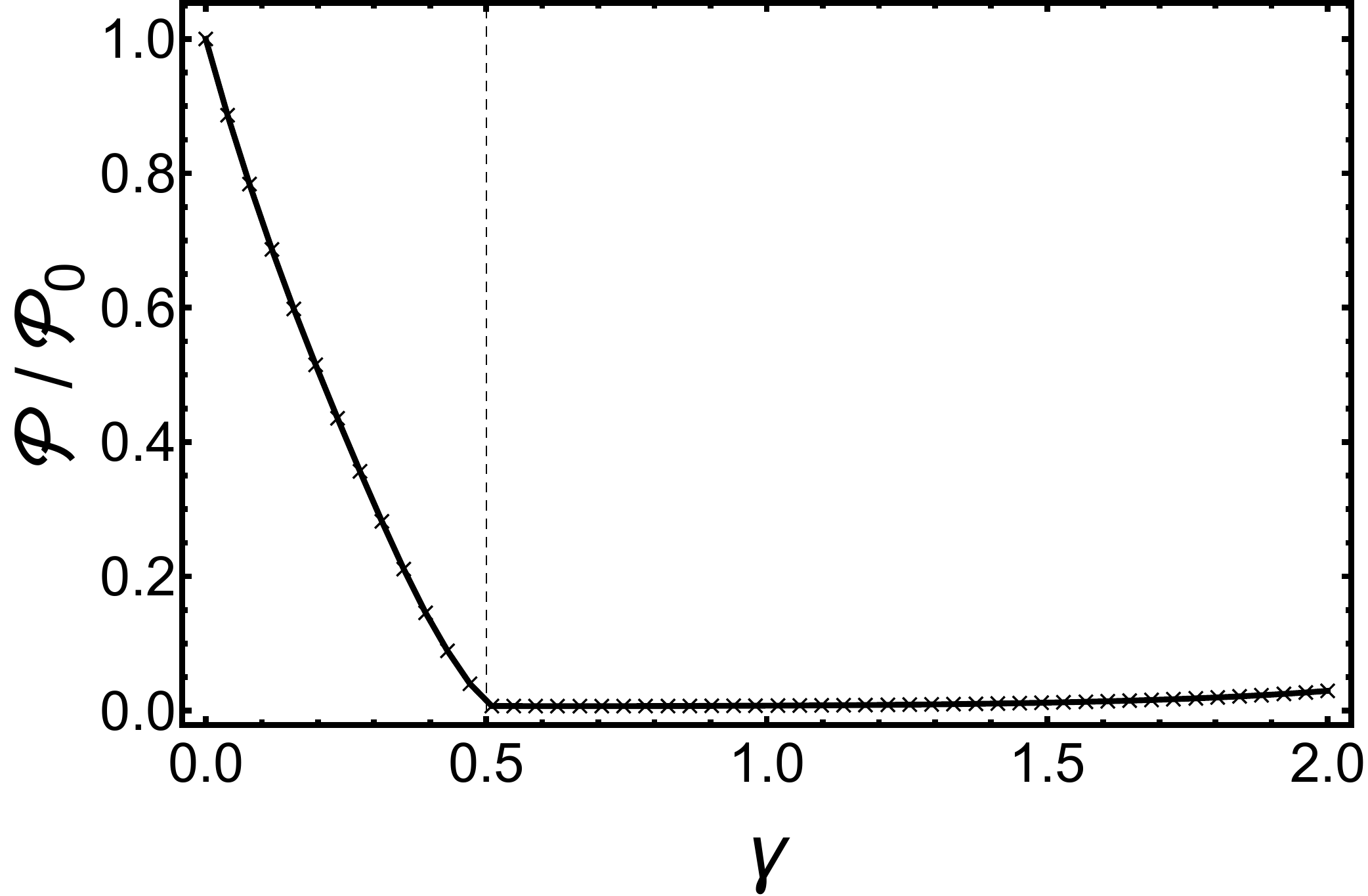}
\includegraphics[width=0.49\columnwidth]{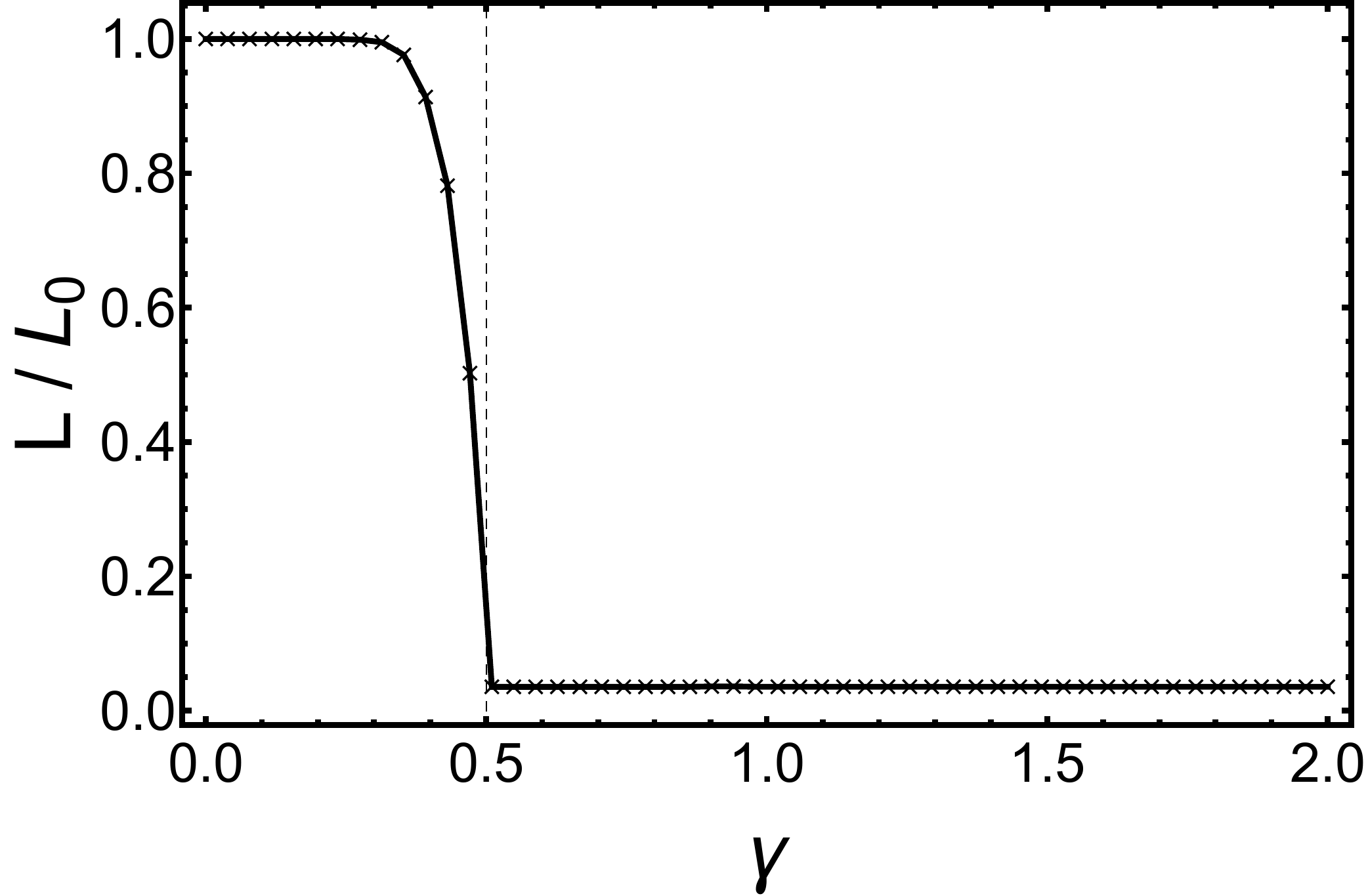}
 \caption{Normalised dissipated energy per unit time at steady state ${\cal P}/{\cal P}_0$ and normalised graph length $L/L_0$ of the steady states, 
as a function of $\gamma$, for the initial network of Figure~\ref{fig1}a).  The initial dissipated energy and the initial length of the graph length are ${\cal P}_0$ and $L_0$, respectively. For the initial conditions of Figure~\ref{fig1}b), the corresponding graphs  are similar.}
\label{fig4}
\end{figure}

In Figure~\ref{fig4}, we show the  dependence on $\gamma$ of these two quantities characterising the phase transition.   The transition observed at $\gamma=1/2$ suggests the existence of a discontinuity on the first derivative of the dissipated energy per unit time \eqref{eq5}, calculated at steady state, relative to the order parameter $\gamma$, which may correspond, in the Ehrenfest classification,  to a first order phase transition. Further simulations show that the phase transition is independent of the initial distribution of conductivities, of the distribution of sources and sinks and of the positions of the nodes of the mesh. Numerical analysis shows that  ${\cal P}(\gamma)$ has a local minimum for $\gamma=2/3$,  as derived in \eqref{eq10}.

\begin{figure}[htb]
\centering
\includegraphics[width=0.49\columnwidth]{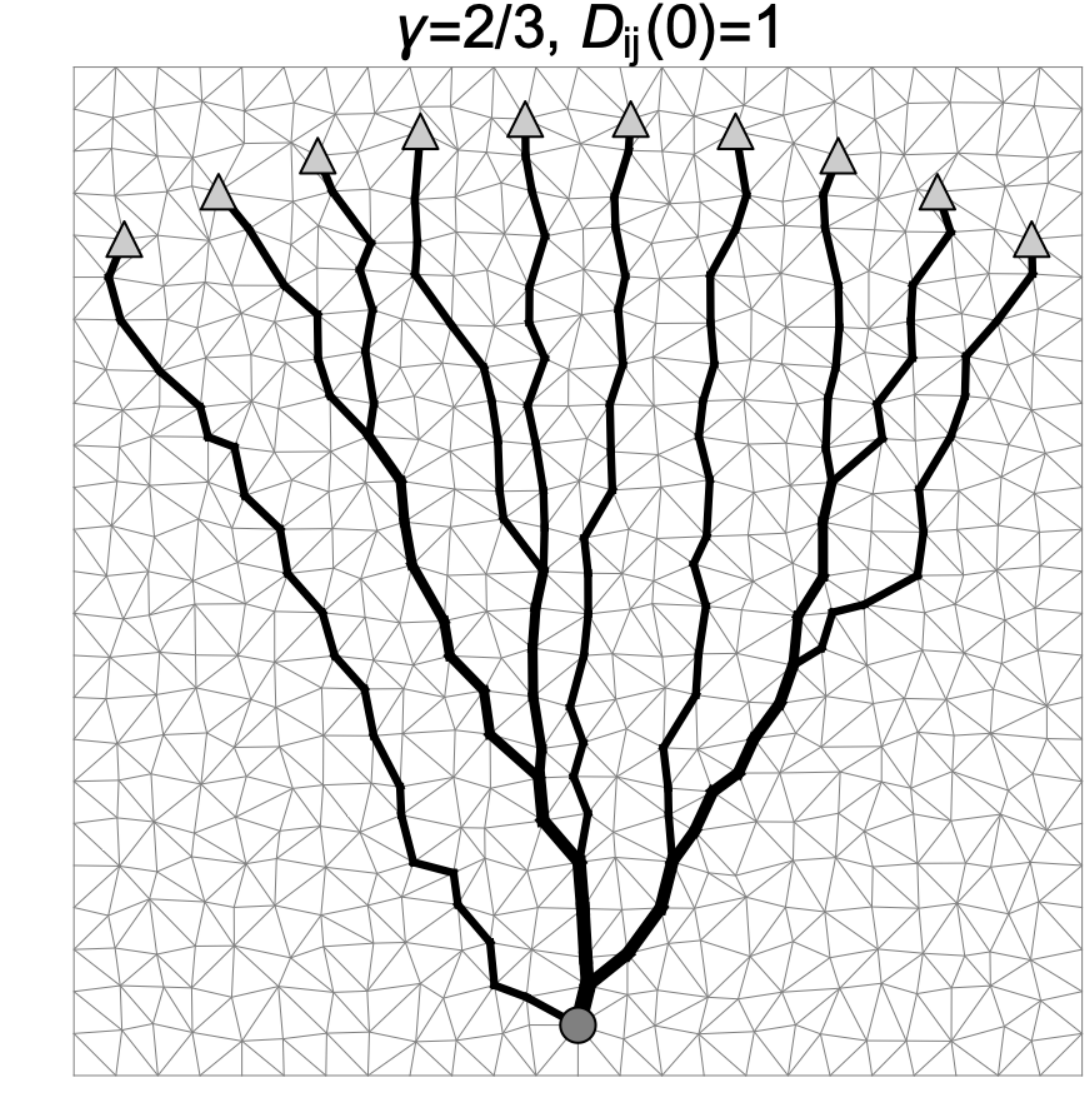}
\includegraphics[width=0.49\columnwidth]{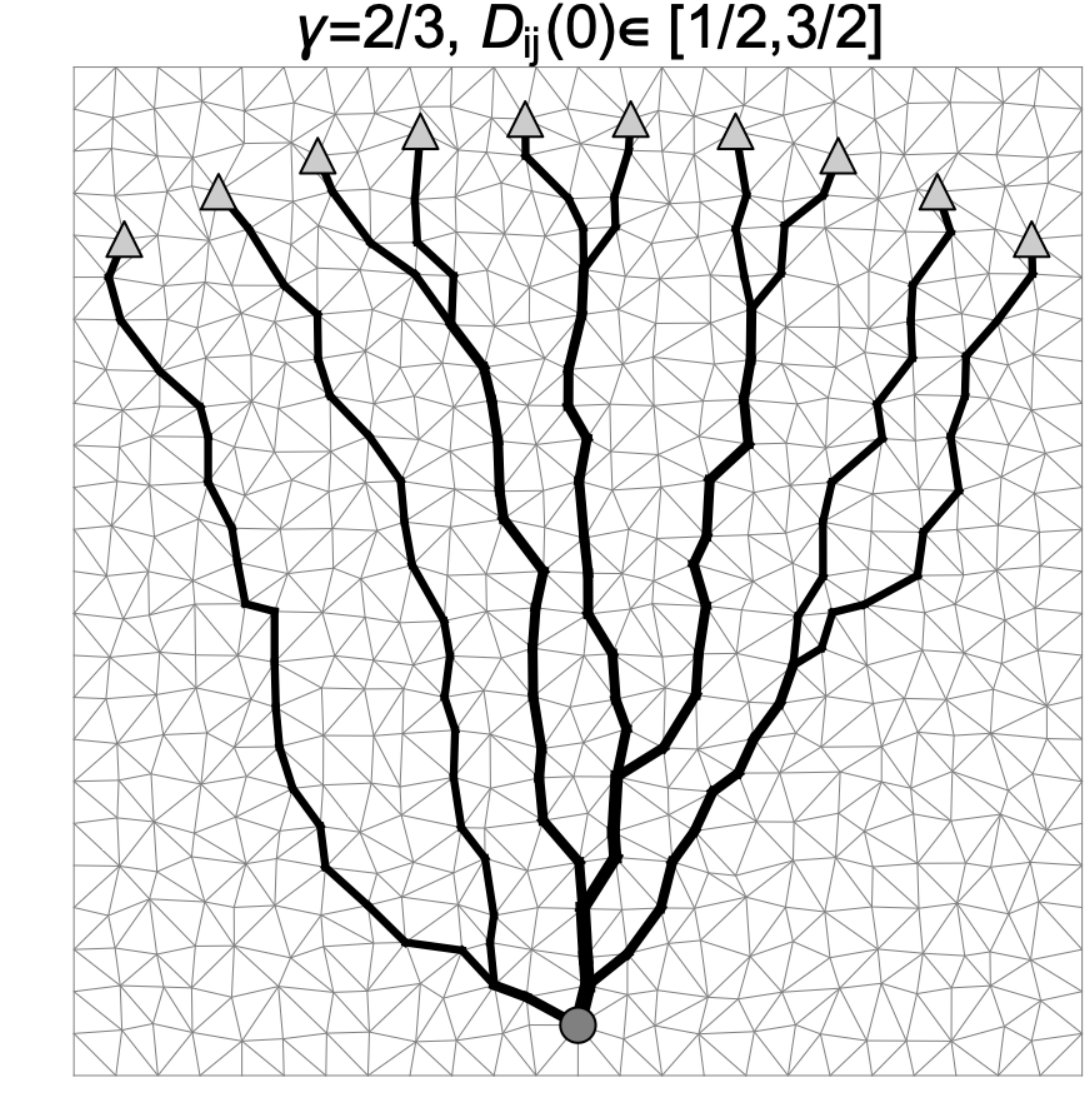}
 \caption{Steady state of the adaptive H-P flow with one source (circle) and ten sinks (triangles), for $\gamma=2/3$,  and initial conductivities $D_{ij}(0)=1$ and $D_{ij}(0)\in[1/2,3/2]$, for the same parameters values of the simulations in Figure~\ref{fig1}. For values of $\gamma$ in the interval $(0.5,2]$, the results are similar. The intensity of the source is $S_{source}=1$, and the intensities of the sinks are $S_{sink}=-1/10$. The thickness of the black lines is proportional to the radius  of  the  edges.}
\label{fig5}
\end{figure}

To  establish further the analogy between the H-P flow and the flows in biological and physical systems, we have simulated an H-P adaptive flow with a geometry similar to the ones found in plant leaves, tumor vascularisation or electric circuits. The steady states of the adapted H-P flows are shown in Figure~\ref{fig5}. In these simulations, the steady stated are different but are metrically close. The structure of the tree network is similar to the vascularisation channels found in plant leaves \cite{Kat}, or electric circuits \cite{Bon}. However, is all these simulations, we have not found  edge loops  in directions transverse to the radial flow emanating from food sources, as observed in \textit{Physarum}.

For graphs describing mazes, the adaptation algorithm described here converges to a path connecting the entrance with the goal of the maze, a phenomenon reported in \textit{Physarum} \cite{Nak}. 

Some of the results described here are based on extensive numerical simulations and on the convergence criteria assumed for the conductivities or graph weights. The phase transition, the steady state trees, the discontinuity of the derivative of the dissipated energy per unit time and the discontinuity of the derivative of the graph length relative to the order parameter $\gamma$ may be derived exactly in the limit when the areas of the Delaunay triangles goes to zero. In this case, the steady state conductivities limits, $D_{ij}^*$ and $D_{ij}^{**}$,  should also be reached.

\textit{Conclusions and outlook.}--  We have derived a class of equations describing an adaptive H-P fluid flow in networks of straight channels, embedded in  any Euclidean space, independently of the dimension. The adaptation mechanism models the transverse elasticity of channels. The flow preserves the volume of the fluid in the network and the channel diameters converge to fixed values, minimising the dissipated energy per unit time. This class of equations have a very general form, enabling the introduction of different choices for the adaptive elasticity functions of channels, simulating different biological and physical systems. In general, this adaptation formalism can be extended to networks with time- and space-dependent sources and sinks  and different mesh topologies.

\end{document}